\documentclass[12pt]{iopart}

\usepackage[utf8]{inputenc}
\usepackage[T1]{fontenc}
\usepackage[english]{babel}
\usepackage[normalem]{ulem}

\usepackage{graphicx}
\usepackage{xcolor}

\usepackage{iopams}
\usepackage{amstext}
\usepackage{amssymb}

\usepackage[colorlinks=true, allcolors=blue]{hyperref}
\usepackage[all]{hypcap}  

\renewcommand{\v}[1]{\boldsymbol{#1}}
\newcommand{\vu}[1]{\hat{\boldsymbol{#1}}}
\newcommand{\dyad}[1]{|#1\rangle\langle#1|}
\newcommand{\Rubidium}{$^{87}\textrm{Rb}$}

\newcommand{\braket}[2]{\langle #1 | #2 \rangle}

\newcommand{\ket}[1]{\left| #1 \right\rangle}
\newcommand{\avg}[1]{\left\langle #1 \right\rangle}

\begin{document}
\title[An ultracold heavy Rydberg system]{An ultracold heavy Rydberg system formed from ultra-long-range molecules bound in a stairwell potential}
\date{\today}
\author{F Hummel$^{1,2}$, P Schmelcher$^{1,3}$, H Ott$^4$, H R Sadeghpour$^2$}
\ead{frederic.hummel@physnet.uni-hamburg.de}
\address{$^1$ Zentrum für Optische Quantentechnologien, Fachbereich Physik, Universität Hamburg, Luruper Chaussee 149, 22761 Hamburg, Germany}
\address{$^2$ ITAMP, Harvard-Smithsonian Center for Astrophysics 60 Garden St., Cambridge, Massachusetts 02138, USA}
\address{$^3$ The Hamburg Centre for Ultrafast Imaging, Universität Hamburg, Luruper Chaussee 149, 22761 Hamburg, Germany}
\address{$^4$ Research Center OPTIMAS, Technische Universität Kaiserslautern, 67663 Kaiserslautern, Germany}

\begin{abstract}
We propose a scheme to realize a \textit{heavy Rydberg system} (HRS), a bound pair of oppositely charged ions, from a gas of ultracold atoms. 
The intermediate step to achieve large internuclear separations is the creation of a unique class of ultra-long-range Rydberg molecules bound in a stairwell potential energy curve.
Here, a ground-state atom is bound to a Rydberg atom in an oscillatory potential emerging due to attractive singlet $p$-wave electron scattering.
The utility of our approach originates in the large electronic dipole transition element between the Rydberg- and the ionic molecule, while the nuclear configuration of the ultracold gas is preserved.
The Rabi coupling between the Rydberg molecule and the heavy Rydberg system is typically in the MHz range and the permanent electric dipole moments of the HRS can be as large as one kilo-Debye.
We identify specific transitions which place the creation of the heavy Rydberg system within immediate reach of experimental realization.
\end{abstract}



\maketitle

\section{Introduction}

A bound pair of oppositely charged ions can be viewed as the molecular analogue of an atomic Rydberg state in which the electron is replaced by the negatively charged anion, hence termed a heavy Rydberg system (HRS) \cite{Kirrander2013, Reinhold2005}. 
Such HRS do form in ionic molecules, such as alkali-halides \cite{Cornett1999}, and whose vibrational energy spectra follow the Rydberg quantum defect formula, 
\begin{equation}
    E_\nu = -\frac{R_\infty}{\mu(\nu-\delta)^2}, \label{eq:ryd}
\end{equation}
in atomic units, where $E_\nu$ is the energy of the vibrational level $\nu$, $R_\infty$ is the Rydberg constant, and $\delta$ is the quantum defect. In electronic Rydberg systems, the reduced mass $\mu\approx m_e$, where $m_e$ is the electron mass, while for the homonuclear HRS, $\mu=M/2$, where $M$ is the atomic mass.

Reactions of anions and cations, leading to neutralization, are important for the ionization balance of various cold astrophysical plasmas \cite{Harada2008}. HRS are typically created from tightly bound molecules \cite{Reinhold2005, Vieitez2008, Beyer2018, Ban2004} and long-range HRS have only been populated via excitation of complex resonances, which are required due to vanishing Franck-Condon overlaps with the vibrational states \cite{Jungen1981}. Excitation of HRS in an ultracold gas, which automatically provides the desired inter-atomic distances, was proposed recently via magnetic Feshbach resonances \cite{Kirrander2013} or with field-control to couple covalent and ion-pair potentials \cite{Markson2016}. 
Formation of HRS via coupling to covalent Rydberg states in the long range in an ultracold environment has several advantages: because these states may form away from the complicated short-range molecular interaction region, there's better control over their formation and spectroscopy; the decay channels which mainly consist of mutual neutralization into atomic species is greatly suppressed \cite{Launoy2019}; and the possibility that the ion pair constituents are both at ultracold temperatures and therefore may lead to formation of a two-component strongly coupled plasma.
Ultracold HRS may eventually serve as a source for production of ultracold negative ions, which are not amenable to laser cooling due to the fact that negative ions have typically a single bound state and hence no cooling transition exists. So far only few atomic species have been identified that offer prospects to cool them \cite{Walter2014,Tang2019}.

In this work, we propose to utilize the excitation of a new class of ultra-long-range Rydberg molecules (which for brevity we refer to as Rydberg molecules from now on) as intermediate states for the preparation of the HRS states. The binding mechanism in a Rydberg molecule is mediated by the repeated scattering of a nearly-free Rydberg electron from a perturber ground-state atom  \cite{Greene2000,Bendkowsky2009,Shaffer2018,Fey2019}. The resulting molecular potentials capture the oscillatory character of the Rydberg electronic wave functions. 

\begin{figure}
\centering
\includegraphics[width=0.5\textwidth]{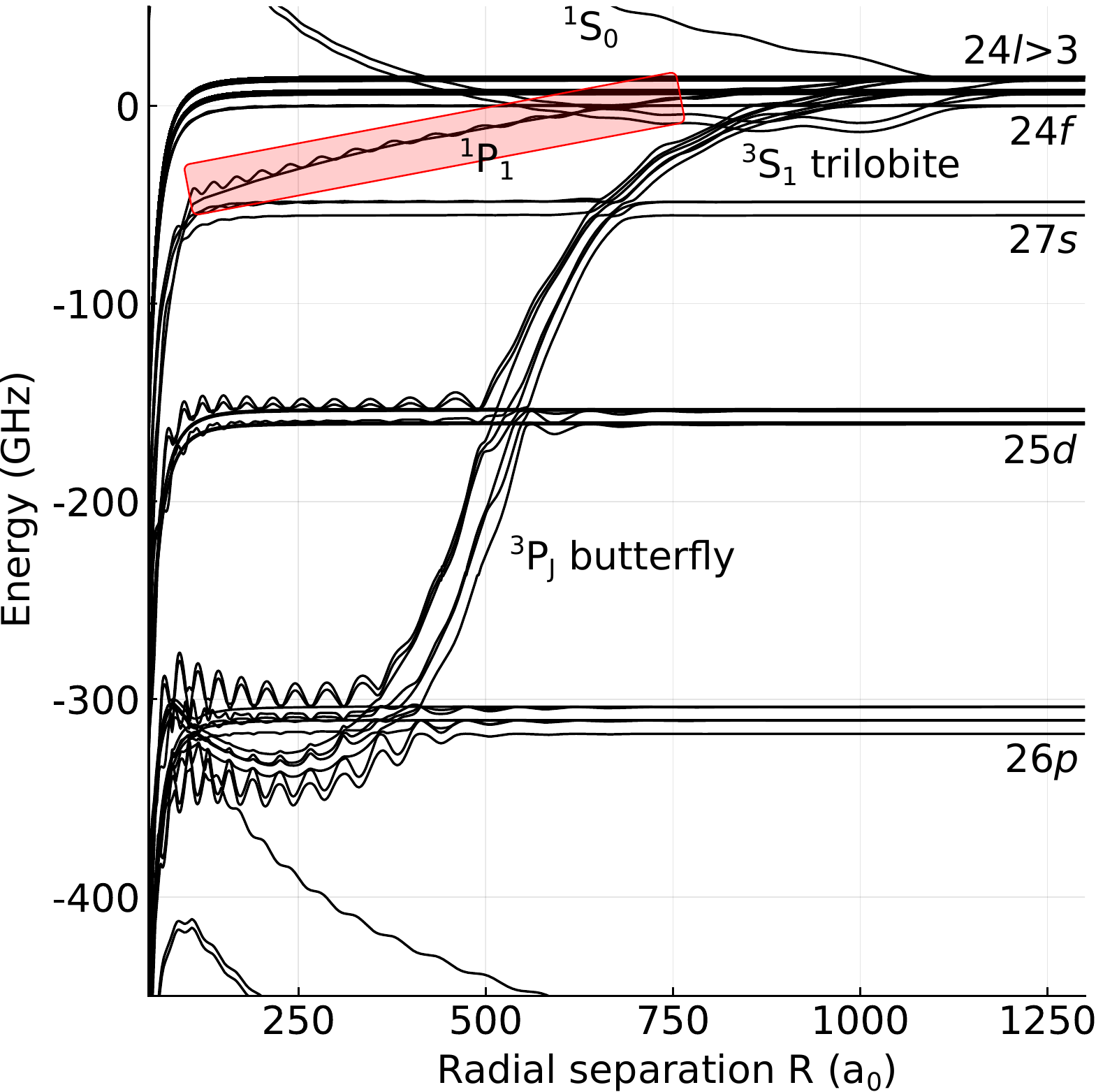}
	\caption{Overview of the molecular potentials near the  Rb($n=24$) Rydberg manifold for $\Omega=\frac{3}{2}$, including the quantum-defect states. Typical potential shapes can be associated with the available scattering channels: the trilobite curves emerging from the quasi-degenerate manifold of $l>3$-states with predominantly $S$-wave triplet scattering (${}^3S_1$) character, the steep butterfly curves crossing through the quantum defect thresholds with triplet $P$-wave scattering (${}^3P_J$) character, a blue-detuned curve with repulsive singlet $S$-wave scattering (${}^1S_0$) character, and a stairwell curve for singlet $P$-wave scattering (${}^1P_1$) highlighted by the shaded area. \label{fig:scheme}}
\end{figure}

The electronic state of the anion-cation pair has the $^1\Sigma$ singlet molecular symmetry for the alkali-metal cases. For the formation of URLM in a magneto-optical trap, it is possible to select different molecular symmetries, due to spin-dependent scattering of the Rydberg electron off the perturber \cite{Khuskivadze2002,Markson2016a,Eiles2017}, which was also experimentally demonstrated \cite{Anderson2014,Sassmannshausen2015,Boettcher2016,MacLennan2019,Engel2019,Deiss2019}. To maximize the transition strength to the anionic ${}^1S_0$ state, a Rydberg molecular state has to be prepared that provides two properties: First, electronic density in the vicinity of the perturber. Second, electronic singlet $P$-wave ($^1P_1$) character with respect to the perturber. To this end, we investigate a \emph{stairwell} Rydberg molecular potential, that has not previously been studied and represents the singlet analogue of the butterfly potentials \cite{Chibisov2002,Hamilton2002,Niederprum2016} (cf. Fig. \ref{fig:scheme}). Similar to the butterfly potentials, the electronic state corresponding to Rydberg molecules formed in a well of this potential can be expressed as a superposition of high-angular momentum Rydberg wave functions. In order to gain spectroscopic access to this type of molecule, we exploit the fact that in \Rubidium, the stairwell potential crosses and mixes with the Rydberg $f$-state, which is detuned slightly from the hydrogenic manifold due to a small quantum defect. Thus, molecules can be excited via a three-photon transition.

Specifically, we demonstrate the charge transfer process by exciting a Rydberg molecule red detuned off the \Rubidium($24f$) Rydberg state, which is formed on the stairwell potential due to attractive $^1P_1$ scattering. An infra-red photon can then be used to drive a transition to the anionic ${}^1S_0$ state. The resulting HRS molecular state is the vibrational analogue of highly excited, isotropic Rydberg electronic states and can be probed spectroscopically over a wide range of HRS principal quantum numbers $5120\lesssim\nu\lesssim5250$.

Before we discuss the results that provide specific experimental requirements for the realization of the proposed excitation scheme in section \ref{sec:results}, in the following section \ref{sec:methods}, we introduce the methods used to theoretically derive the Rydberg molecular potential energy curves (PECs) and the vibrational structure of Rydberg molecules. Furthermore, we introduce the model for the HRS. Section \ref{sec:conc} provides the conclusion.

\section{Methodology  \label{sec:methods}}

\subsection{Ultra-long-range Rydberg molecule} 

Let us consider two rubidium atoms, one of which is excited to a Rydberg state. In the spirit of the Born-Oppenheimer approximation, we separate the nuclear from the electronic degrees of freedom, while the adiabatic electronic eigenstates and corresponding energies depend parametrically on the internuclear separation $R$. Solving the time-independent Schrödinger equation (SE) for the electronic problem provides the PECs which are used as input for the vibrational SE.
We employ the electronic Hamiltonian (in atomic units) \cite{Eiles2017}
\begin{equation}
    H_\text{e} = H_\text{R} + H_\text{G} + V_\text{p} + V_\text{e}. \label{eq:elham}
\end{equation}
$H_\text{R}$ describes the dynamics of the Rydberg electron at position $\v{r}$ in the potential of the ionic core at the origin. The electron has spin $\v{s}_1$ and orbital $\v{l}$ angular momentum. The eigenstates of $H_\text{R}$, $\phi_{nljm_j}(\v{r})$, depend on the principal quantum number $n$, orbital angular momentum $l$, and the total angular momentum $j=|\v{s}_1+\v{l}|$ with its projection onto the internuclear axis $m_j$. The eigenvalues are taken from spectroscopic measurements \cite{Li2003, Han2006} and are used to analytically express the long-range behavior of $\phi_{nljm_j}(\v{r})$ in terms of Whitaker functions.

The ground-state atom is considered in terms of its hyperfine interaction $H_\text{G}=a\v{I}\cdot\v{s}_2$, where $\v{I}$ is the nuclear spin, $\v{s}_2$ is the spin of the valence electron, and $a$ is the hyperfine constant taken from \cite{Arimondo1977}. Eigenstates of $H_\text{G}$ are $\ket{F,m_F}$, where $F=|\v{I}+\v{s}_2|$.
The ionic Rydberg core polarizes the ground-state atom, which is considered by $V_\text{p}=-\frac{\alpha}{2R^4}$, where the polarizability $\alpha$ is taken from \cite{Mitroy2010}. 

The interaction of the Rydberg electron and the ground-state atom is modeled by a Fermi-type pseudopotential \cite{Fermi1934}, which has been generalized to include orbital angular momentum $\v{L}$ of the Rydberg electron relative to the ground-state atom \cite{Omont1977} up to $P$-wave interaction ($L=1$; $L=0$ corresponds to $S$-wave interaction) as well as the total electronic spin of the Rydberg electron and the valence electron $\v{S}=\v{s}_1+\v{s}_2$ in terms of singlet ($S=0$) and triplet ($S=1$) scattering channels.
For this contact interaction, we employ the potential \cite{Eiles2017}
\begin{equation}
    V_\text{e} = \sum_\beta a_{SLJ}(k) \frac{(2L+1)^2}{2x^{2(L+1)}} \delta(x) \dyad{\beta}, \label{eq:scatpot}
\end{equation}
where $x=|\v{r}-\v{R}|$ is the distance between the Rydberg electron and the ground-state atom and the sum is taken over available interaction channels defined by the multi index $\beta$, where $\ket{\beta}=\ket{(LS)JM_J}$. Here, $J=|\v{L}+\v{S}|$ is the total angular momentum of the two electrons with respect to the ground-state atom's core and $M_J$ its projection onto the internuclear axis. The spin- and orbital angular momentum dependent scattering lengths and volumes $a_{SLJ}(k)$ are derived from the respective phase shifts, which are taken from \cite{Engel2019}. They are energy dependent and the wave number $k$ is calculated via the semi-classical relation $k=\sqrt{2/R-2E^\star}$, where $E^\star$ is the asymptotic atomic energy of \Rubidium($24f$).
The only good quantum number of the Rydberg molecule is the projection of the total angular momentum onto the internuclear axis $\Omega = m_j + m_F$. 

Without loss of generality, the internuclear axis is assumed to be the $z$ axis. Solutions to the electronic problem and the PECs are obtained by diagonalizing $H_\text{e}$ (\ref{eq:elham}) in a finite basis that includes two hydrogenic manifolds, one of which lies energetically above the state of interest $24f$ ($n=24$) and one lies below ($n=23$). All angular momenta $j$ are considered, while projections $|m_j|>\frac{3}{2}$ are neglected since they do not interact with the ground-state atom. The basis $\ket{Fm_F}$ is included completely.
The chosen basis set has proven to be sufficiently accurate despite the inherent convergence issues that stem from the non-local character of the contact scattering interaction \cite{Fey2015, Engel2019}.

After solving the electronic problem, a specific molecular potential $V_\text{in}(R)$ is used as input for the nuclear vibrational SE which is solved by standard one-dimensional finite difference methods. 
Since the energy spacing of vibrational states is on the order of few GHz, we neglect couplings to rotational degrees of freedom, the energetic scale of which is three orders of magnitude smaller for the given internuclear separations.

The vibrational ground-state in the given PEC $V_\text{in}(R)$ is termed initial state $\chi_\text{in}(R)$ for the creation of the heavy Rydberg system. This type of state is typically a Gaussian-like wave packet without nodes. Its center of mass is given by $R_\text{in} = \braket{\chi_{in}|R}{\chi_\text{in}}$.
The corresponding electronic wave function for this equilibrium position is termed $\psi_\text{in}(\v{r};R_\text{in})=\braket{\v{r}}{\psi_\text{in}}$. 
The electronic spin character of the initial state is obtained by projecting the electronic wave function 
$\sigma_{S}(R_\text{in}) = |\braket{S}{\psi_\text{in}}|^2$.

\subsection{Heavy Rydberg system}

To model the heavy Rydberg system, we employ the ionic potential
\begin{equation}
    V_\text{f} = -E_{ea} - \frac{1}{R} - \frac{\alpha_p + \alpha_m}{2R^4}, \label{eq:ionpot}
\end{equation}
where $E_{ea}=117.5\,\text{THz}$ is the electron affinity of rubidium \cite{Frey1978} and $\alpha_{p/m}=9.11/526\,a_0^3$ is the polarizability of the positively/negatively charged rubidium ion \cite{Mitroy2010, Fabrikant1993}.
In general, the vibrational nuclear wave function can be expressed in terms of the hydrogenic wave functions $\ket{\chi_\nu^\Lambda}$ with the appropriate reduced mass $\mu$ and with principal quantum number $\nu$ and angular momentum $\Lambda$, which correspond to vibrations and rotations, respectively:
\begin{equation}
    \ket{\chi_\text{f}} = \sum_{\nu \Lambda} c_{\nu \Lambda} \ket{\chi_\nu^\Lambda}. \label{eq:final}
\end{equation}
Here, up to normalization, the coefficient $c_{\nu \Lambda}$ equals the nuclear Franck-Condon overlap $\braket{\chi_\nu^\Lambda}{\chi_\text{in}}$ of a specific vibrational-rotational state and the Rydberg molecular initial state $\ket{\chi_\text{in}}$. In our case, the range of available $c_{\nu \Lambda}$ depends on the spectroscopic width of the transition from the Rydberg molecule to the HRS. Typically, by aid of a narrow-band laser, each level $\nu$, the energy of which is given by equation (\ref{eq:ryd}), can be addressed individually and the sum over $\nu$ in equation (\ref{eq:final}) collapses to a specific value $\nu'$.

For the anion electronic wave function, we employ an Eckart wave function \cite{Eckart1930} with the functional form 
\begin{equation}
    \braket{\v{r}}{\psi_\text{f}} = N \cosh^{-2\lambda}\left(\frac{x}{r_0}\right) \sinh\left(\frac{x}{r_0}\right),
\end{equation}
where $N$ is a normalization constant, $x=|\v{r}-R_\text{in} \vu{e}_z|$ is the distance of the electron to the position of the anion nucleus $R_\text{in}$, and $\lambda=1.35094$ and $r_0=9.004786\,a_0$ have been obtained to reproduce the electron affinity and the electron-atom $S$-wave scattering length of \Rubidium\,\cite{Markson2016}.

\section{Results and discussions \label{sec:results}}

\subsection{\it{Stairwell} Rydberg molecular potential curve}

\begin{figure}
\centering
\includegraphics[width=0.5\textwidth]{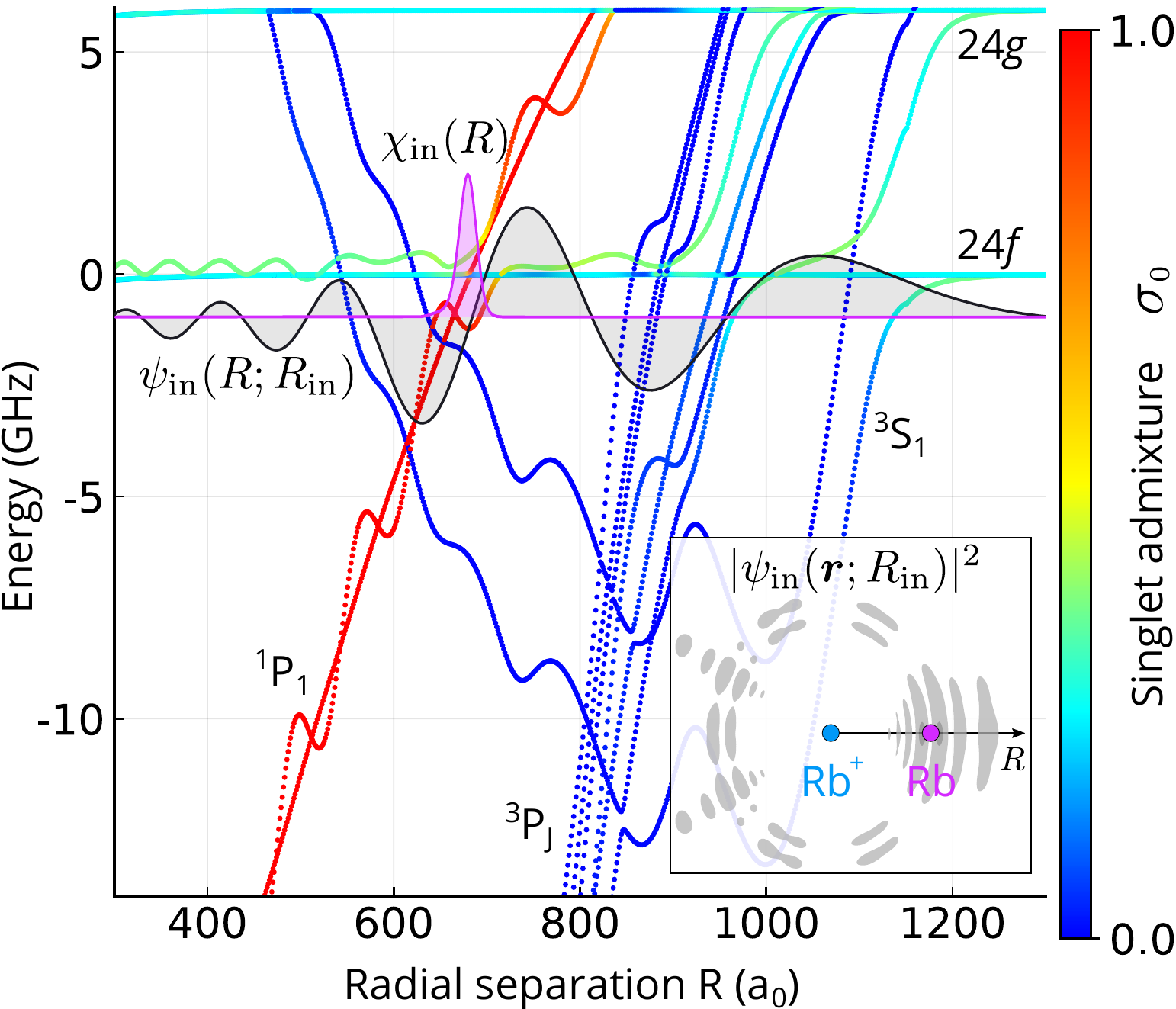}
	\caption{Molecular potentials correlated to the \Rubidium($24f$) Rydberg state for $\Omega=\frac{3}{2}$. The color bar indicates the degree of singlet mixing character in the molecular potentials. Trilobite curves with outer minima around $R\sim 1000 \text{a}_0$ are visible in blue (${}^3S_1$) and crossed around $R=800\,a_0$ by steep butterfly curves (${}^3P_J$). In the potential well around $R_\text{in}=678\,a_0$, a vibrational state can be formed  which serves as initial state $\chi_\text{in}(R)$ (purple). A radial cut of the corresponding electronic wave function $\psi_\text{in}(R;R_\text{in})$ along the internuclear coordinate, but for the ground-state atom being fixed at $R=R_\text{in}$, is shown in black. Its node at the potential minimum reflects the $P$-wave character relative to the ground-state atom and its singlet admixture is $\sigma_{0}=0.94$. A contour of the electronic density is shown in the inset. \label{fig:ULRMpot}}
\end{figure}

To calculate the transition from the Rydberg molecule to the HRS, we determine the initial and final states of the respective systems, individually. The ultra-long-range dimer consists of a single \Rubidium($5s$) ground-state atom inside the orbit of a \Rubidium($24f$)-Rydberg state. In rubidium and other alkali metals, the $f$ states have a small quantum defect $\delta_f=0.01637$ \cite{Han2006}.

The molecular potentials in the region of the atomic $n=24$-Rydberg state in the $\Omega=\frac{3}{2}$ molecular symmetry are given in Fig.\,\ref{fig:scheme}. At large internuclear distances, the potentials are flat and the energies reflect the atomic Rydberg energies. The multiplets of the quantum defect states arise from the Rydberg fine structure and ground-state hyperfine structure splittings.
The zero energy on the vertical axes is set to the asymptotic atomic energy of the Rb($24f_{5/2}$) with the ground-state atom in a hyperfine state $F=1$.
The archetypal trilobite \cite{Booth2015} and butterfly \cite{Niederprum2016} molecules form potentials which detach from the hydrogenic manifold. Trilobite molecules form due to the attractive $S$-wave scattering length \cite{Greene2000}, while butterfly molecules form due to resonant $P$-wave scattering \cite{Hamilton2002, Chibisov2002}. In both cases, the corresponding total electronic state of the two participating valence electrons is a triplet state, hence the labeling ${}^3S_1$ and ${}^3P_J$, respectively. Additionally, a different potential emerges from the degenerate manifold that corresponds also to the $P$-wave scattering, but in the electronic singlet channel (${}^1P_1$), here highlighted by the red box, and hereafter called the \emph{stairwell} potential. This potential crosses the $f$ electronic state, which in turn can be employed to access the molecular states.
The stairwell potential is also present for $\Omega=\frac{1}{2}$, while for $\Omega=\frac{5}{2}$, only the non-oscillatory potential that does not support bound states remains. 

A more detailed presentation of the PECs is shown in Fig.\,\ref{fig:ULRMpot}. The singlet character of the potentials is represented in color; red curves have mainly singlet, while blue curves have mainly triplet character. 
The ${}^1P_1$ curve has an avoided crossing with the $f$ state at around $R=700\,a_0$ below which a potential well is visible. It supports vibrational states, the ground-state of which is visible in Fig.\,\ref{fig:ULRMpot} as the purple curve and is labeled $\chi_\text{in}(R)$, for it serves as the initial vibrational state for the transition to the HRS. Its mean radial position and the minimum of the potential well lie at $R_\text{in}=678\,a_0$.

A radial cut of the electronic wave function $\psi_\text{in}(R;R_\text{in})$ along the internuclear coordinate $\v{r}=R\vu{e}_z$ corresponding to the fixed internuclear distance $R_\text{in}$ is shown to scale as the black curve, for it serves as the initial electronic state for the transition to the HRS. Its $P$-wave character relative to the ground-state atom is reflected by the node at the potential minimum, where the gradient of the wave function is maximal. Additionally, the inset shows a contour of the full density of the electronic wave function and the nuclear positions therein. Significant electron density is localized in the vicinity of the ground-state atom, which is also visualized by the maximal amplitude of the wave function next to its node at the position of the ground-state atom.

\subsection{HRS potential curve}

\begin{figure}
	\centering
	\includegraphics[width=0.5\textwidth]{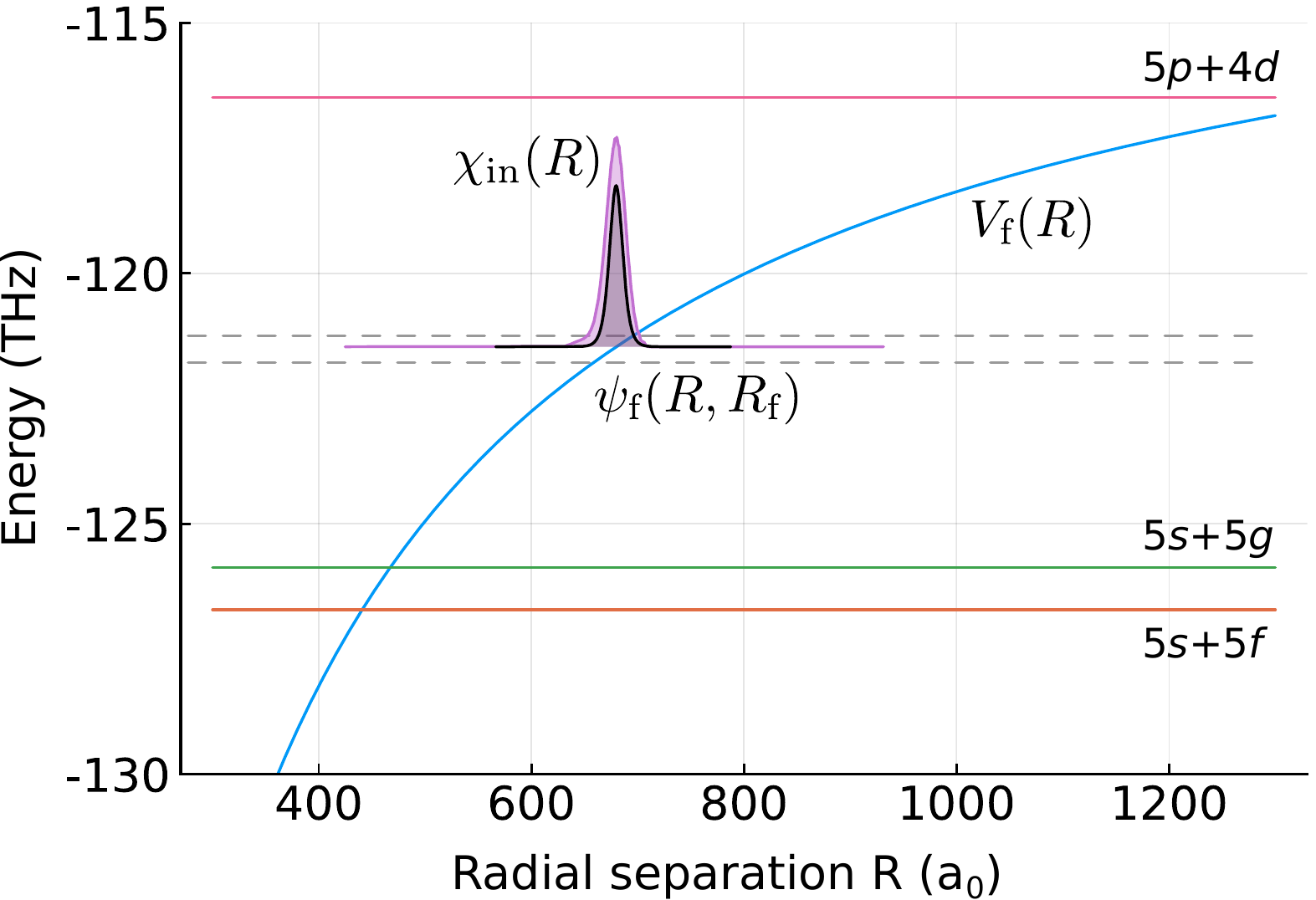}
	\caption{Ion-pair potential dominated by the Coulomb interaction $V_\text{f}(R)$ (blue) and the initial nuclear wave function $\chi_\text{in}(R)$ (purple) along with the final electronic wave function $\psi_\text{f}(R,R_\text{f})$ (black). The grey dashed lines highlight the energy range in which the HRS can be probed. The energy scale is given relative to the $24f$ state similarly as in figure \ref{fig:ULRMpot}. Colored horizontal lines indicate dissociation limits of molecular potentials. \label{fig:hrspot}}
\end{figure}

In the HRS, the excited electron is charge transferred  to the ground-state atom to form a negative ion, which then interacts with the positive ion, through Coulomb and charge-atom polarization potentials. In general, this potential is crossed by other molecular PEC corresponding to electronically low-lying excited molecular states in the dissociation limit. For internuclear separations $R>30\,a_0$ these potentials are flat and lead to narrow avoided crossings with the Coulomb-like potential $V_\text{f}(R)$ \cite{Park2001, Markson2016}. Fig.\,\ref{fig:hrspot} shows the HRS potential $V_\text{f}(R)$ (blue) in the region of interest along with the closest dissociation limits of the \Rubidium\,$5s$+$5f$, $5s$+$5g$, and $5p$+$4d$ states (colored lines). Remote of the crossings, the energy levels of vibrational states in this potential are well described by the Rydberg formula given in Eq.\,(\ref{eq:ryd}). 
As a corollary to Coulombic three-body recombination, the mechanism which "recombines" the ions here is called the mutual neutralization.

The anion electronic wave function localizes at the initial position of the ground-state atom $R_\text{in}$. A radial cut $\psi_\text{f}(R,R_\text{f})$ is shown in Fig.\,\ref{fig:hrspot} as black curve while the initial vibrational wave function $\chi_\text{in}(R)$ is shown in purple.

When the ultracold gas is prepared experimentally, we expect it to be homogeneous and isotropic. The electronic Rydberg transition to prepare the initial Rydberg molecule does not alter this isotropic nuclear configuration: Due to the absence of external fields, no angular dependence is introduced by the Rydberg molecular potentials and the dimer depends only on the radial internuclear coordinate. The Franck-Condon transition from Rydberg molecule to HRS still maintains the isotropy and consequently only isotropic states $\ket{\chi_\nu^0}$ contribute to the overlaps. From the given initial state $\ket{\chi_\text{in}}$, the range of available energy levels $\nu$ is indicated by the gray dashed lines in Fig.\,\ref{fig:hrspot} (compare discussion in section \ref{sec:FC}).

At $R_\text{in}$, where most of the initial gas atoms are found (largest Franck-Condon factor), the electronic dipole transition element is calculated,
\begin{equation}
    \braket{\psi_\text{f}}{\vu{\epsilon}\v{r}|\psi_\text{in}} = 0.91\,a_0,
\end{equation}
which includes the contribution of the spin-mixing, given as $\braket{S=0}{\psi_\text{in}}=0.96$.
Here, $\vu{\epsilon}$ is the direction of the electric field imposed by the laser, linearly polarized parallel to the internuclear axis. This matrix element is sizeable and on the order of a typical optical dipole transition, allowing for efficient coupling between the Rydberg molecule and the HRS.

\subsection{Optimal Franck-Condon Factors \label{sec:FC}}

\begin{figure}
	\centering
	\includegraphics[width=0.5\textwidth]{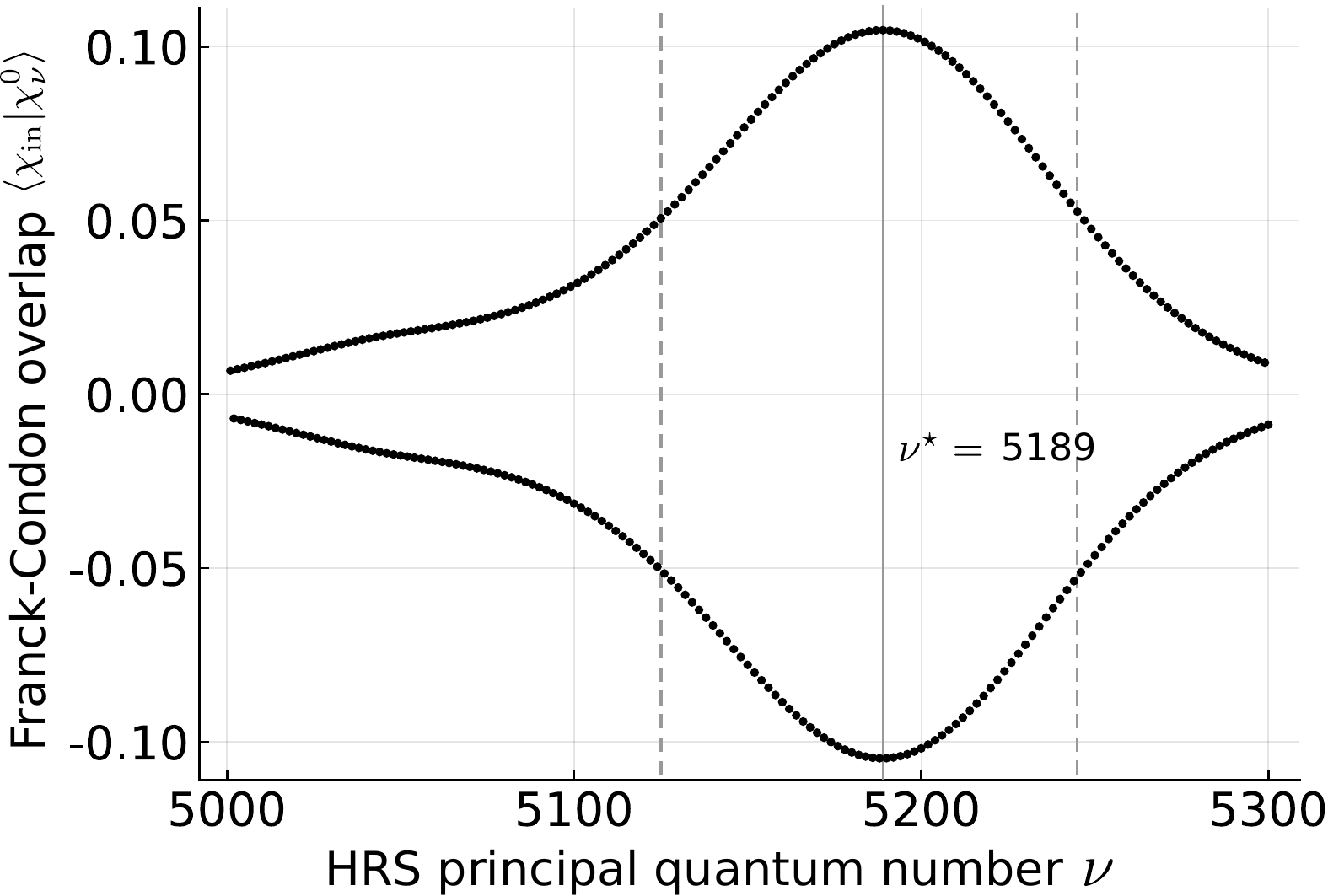}
	\caption{The nuclear Franck-Condon overlap $\braket{\chi_\text{in}}{\chi_\nu^0}$ for the principal quantum number $\nu$ of the hydrogen-type final state with a focus on the dominantly contributing region. Due to the oscillatory nuclear wave function the sign alternates. Maximal overlap is obtained for the isotropic state with $\nu^\star=5189$, indicated by a vertical line. Dashed lines indicate one half of the maximum. On this scale, the respective curves for $\Lambda=\{1,2,3\}$ (not shown) look identical. \label{fig:fc}}
\end{figure}

Due to the spectroscopically resolvable energetic separation between vibrational levels in the HRS, each vibrational state $\ket{\chi_\nu^0}$ can be addressed individually. Figure \ref{fig:fc} shows the distribution of Franck-Condon overlaps $\braket{\chi_\text{in}}{\chi_\nu^0}$ for a range of HRS principal quantum numbers $\nu$. The wave function of the given initial state $\chi_\text{in}(R)$ is shown in Figs.\,\ref{fig:ULRMpot} and \ref{fig:hrspot}. 
The dominant contribution with $\braket{\chi_\text{in}}{\chi_{\nu^\star}^0}=0.105$ is achieved for the vibrational level $\nu^\star=5189$ indicated in Fig.\,\ref{fig:fc} by the vertical line. The energy difference between the initial and this state is $E_{\nu^\star}-E_\text{in}=121.47\,\text{THz}$, which translates to a laser wave length of $\lambda=2468$ nm in the near infra-red. Here, $E_\text{in}$ is the energy of the Rydberg molecular vibrational ground state in the molecular potential $V_\text{in}(R)$ and $E_{\nu^\star}$ is the energy of the HRS according to Eq.\,(\ref{eq:ryd}) omitting the quantum defect $\delta$. Deviations of this energy allow systematic determination of $\delta$ for $\Lambda=0$. We focus on levels where the Franck-Condon overlap exceeds half of the maximal value at $\nu^\star$ indicated in Fig.\,\ref{fig:fc} by the dashed, vertical line. In an energy window of $\sim 500$ GHz, around $E_{\nu^\star}-E_\text{in}$, there are as many as 100 vibrational HRS states. The spacing between individual levels is approximately $3.6$ to $3.9\,\text{GHz}$. For $\nu^\star$, the binding energy in the ionic potential is $9.68\,\text{GHz}$.
If it is possible to excite also rotational states $\Lambda>0$ from the ultracold gas, additional Rydberg series corresponding to each $\Lambda$ will be observed. The distribution of Franck-Condon overlaps for $\Lambda=\{1,2,3\}$ up to seven-digit precision is the same as for $\Lambda=0$, shown in Fig.\,\ref{fig:fc}.

The dipole moment of the HRS can be given analytically $\avg{R}=\braket{\chi_\nu^\Lambda}{R|\chi_\nu^\Lambda} =[3\nu^2-\Lambda(\Lambda+1)]/2\mu$ \cite{Bethe1957} and for $5120\lesssim\nu\lesssim5250$ and $\Lambda=0$ this results in $500\,a_0\lesssim\avg{R}\lesssim520\,a_0$.

In general, also $f$ states with different principal quantum number can be employed as the initial states. We chose the $24f$ state due to the fact that the potential well in which the molecule forms is energetically close to the asymptotic $f$-state Rydberg energy, which guarantees sufficient $f$ admixture to spectroscopically address the molecule by a three-photon transition. Additionally, another class of molecules correlated with the Rb(n$p$) Rydberg atoms can also serve as the initial state, the properties of which vary from the presented ones and are discussed in \ref{sec:app}.

\subsection{Experimental realization \label{sec:exp}}

The experimental generation of a HRS, as outlined above is promising. Both, the excitation of Rydberg $f$ states as well as the coupling of the Rydberg molecule to the HRS can be achieved with the current state-of-the-art laser technology. For rubidium, the photoassociation of the Rydberg molecule requires a three photon scheme, involving, e.g., 776\,nm, 780\,nm and 1280\,nm. All three wavelengths can be realized with tunable diode lasers and an overall three-photon coupling rate in the MHz range can be achieved. Ultracold temperatures as realized by standard laser cooling and high atomic densities of $10^{12}$ cm$^{-3}$ or larger are advisable in order to have a large overlap of the ground state wave function with the Rydberg molecular vibrational wave function.

The second step, the coupling to the HRS, is equally feasible. The transition wavelength of 2468\,nm requires either an OPO system or a DFB diode laser. Thereby, a laser power of a few mW is already sufficient: for a beam waist of 1\,mm, a power of 1\,mW, a Franck-Condon overlap of 0.1 and an electric dipole transition matrix element of 1\,$e a_0$, the Rabi frequency between the Rydberg molecule and the HRS amounts to about 1\,MHz. This time scale is much faster than the lifetime of the Rydberg molecule, which is here only limited by the Rydberg atom, such that the branching ratio is favorable towards the production of a HRS.

The generation of a HRS could be done in a pump-probe type of experiment and the HRS could be detected by photodissociation followed by the detection of the negative ion. The latter can be identified by its time of flight to the detector.

\section{Conclusion \label{sec:conc}}

We show that ultra-long-range Rydberg molecules can be used to produce ultracold heavy Rydberg system i.e.\,a bound anion-cation pair. Due to the large internuclear separations inherent to Rydberg molecules, the resulting HRS can be probed over a wide range of large principal quantum numbers. To our knowledge, HRS have only been produced from tightly bound molecules, in which the Franck-Condon factor strongly constraints possible internuclear separations. In the proposed association scheme, we utilize a molecular potential that has not been studied previously and arises from weakly attractive singlet $P$-wave scattering. The resulting electronic states, similarly to the trilobite and butterfly states, are superpositions of high angular momentum states and, thus, have large densities in the vicinity of the ground-state atom which largely increases the electronic dipole transition element compared to other choices of initial states. Still, studying the plethora of Rydberg molecule potentials is still far from exhaustive and further efforts would be beneficial in order to explore other choices for the initial state. In this spirit, a recent work \cite{Peper2020} has predicted that in cesium also $p$-state molecules with appropriate butterfly admixture provide a suitable initial state.
Creation and study of ultracold HRS will lead to new insights on questions of molecular formation in dilute stellar gases and mutual neutralization processes. Furthermore, HRS serves as the precursor to the creation of equal mass strongly-coupled ultra-cold plasmas.

\ack{P.S. acknowledges financial support from the Deutsche Forschungsgemeinschaft within the priority program "Giant interactions in Rydberg systems" [DFG SPP 1929 GiRyd project SCHM 885/30-1]. This work was financially supported within the PIER Hamburg-MIT/BOS Seed Projects with the Project ID: PHM-2019-06. F.H. is grateful to Matthew Eiles and Kevin Keiler for useful comments and enlightening discussions. }

\appendix
\section{\emph{p}-state molecules \label{sec:app}}

\begin{figure}
\centering
\includegraphics[width=0.45\textwidth]{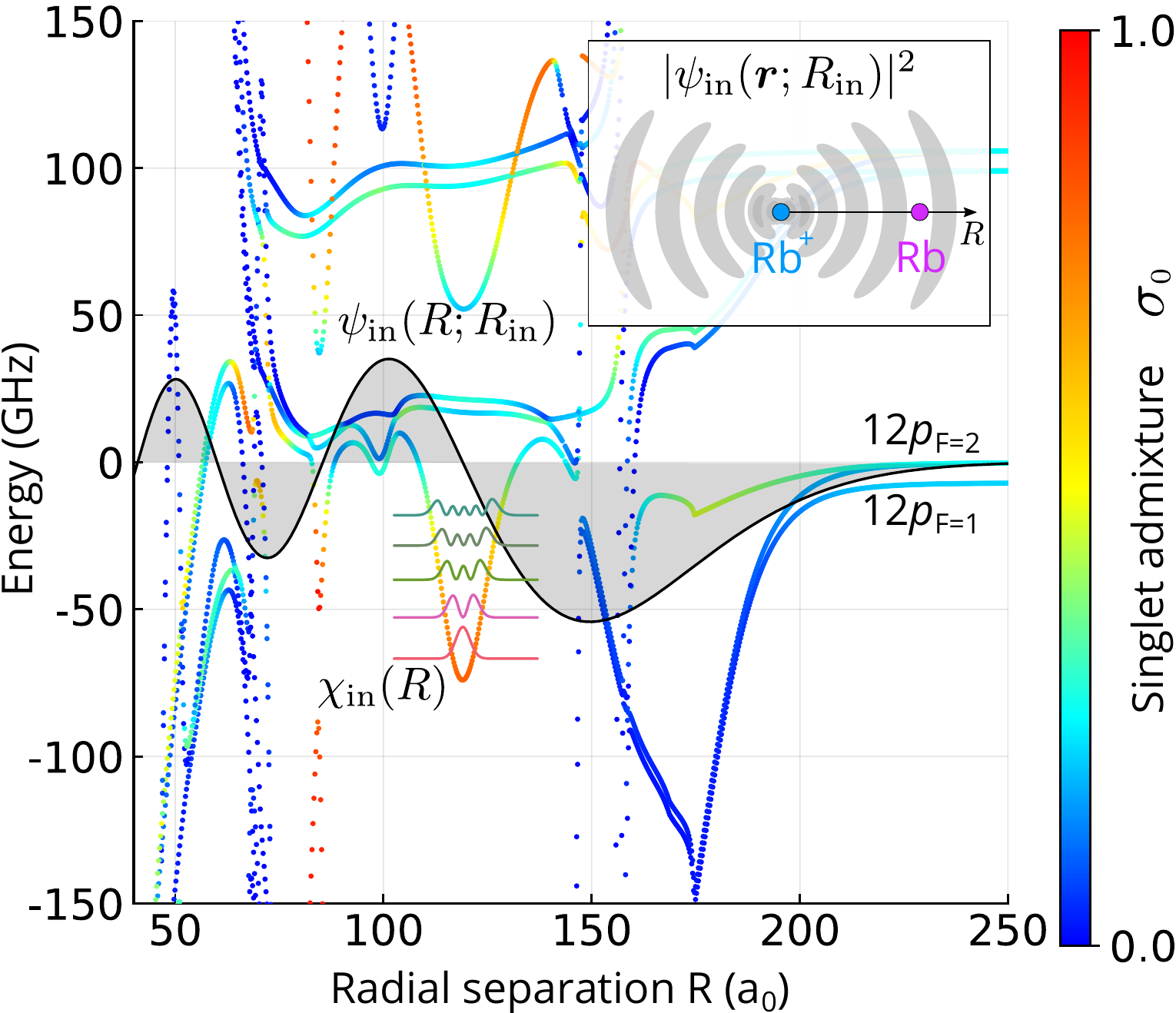}
	\caption{Molecular potentials correlated to the $12p$ Rydberg states for $\Omega=\frac{3}{2}$. The colorcode indicates the electronic singlet character. In the potential well around $R_\text{in}=120\,a_0$, vibrational states can be formed (colored curves), the ground state of which serves as initial state $\chi_\text{in}(R)$. A radial cut of the corresponding electronic wave function $\psi_\text{in}(\v{r};R_\text{in})$ is shown in black. Its node at the potential minimum reflects the $P$-wave character relative to the ground-stat atom and its singlet admixture is $\sigma_{0}\approx0.8$. The inset shows a contour of the electronic density. \label{appfig:ULRMpot}}
\end{figure}

In addition to  Rb($24f$) Rydberg molecular potential energy stairwell curve for initiating the formation of HRS, we present here another example of an initial state, namely the Rb($12p$) Rydberg molecule. The corresponding potentials are far detuned from the hydrogenic manifold (cf. Fig. \ref{fig:scheme}) and the stairwell potential is energetically not available. However, for this relatively low-lying Rydberg state, the semi-classical electron momentum $k$ varies sufficiently quickly along the internuclear separation, that in the outer region of the electron orbit additionally to $S$-wave scattering, $P$-wave channels contribute to the overall shape of the resulting PECs. These potentials can be seen in figure \ref{appfig:ULRMpot}. For large internuclear distances, flat curves are visible representing the atomic Rydberg levels occurring with a slight splitting in energies which depends on the hyperfine level of the ground-state atom. 
For internuclear distances between $150\,a_0<R<200\,a_0$ the atomic energy is lowered due to the presence of attractive triplet $S$-wave scattering which reflects the electronic density. The relative depth of different visible curves arises due to varying singlet admixture of the corresponding electronic state, which is indicated by the colorcode. 
At $R=150\,a_0$ crossings with steep butterfly curves occur, which are induced by a shape resonance in the $P$-wave scattering channel \cite{Hamilton2002, Chibisov2002}. For smaller internuclear distances $100\,a_0<R<150\,a_0$ a potential well with a high singlet admixture is visible. 
The electronic wave function $\psi_\text{in}(R;R_\text{in})$ corresponding to the minimum of the well at $R_\text{in}=120\,a_0$ with $\sigma_0=0.8$ is visible as black curve in the figure (amplitude in arbitrary units) and serves as initial electronic state for the transition to the HRS. Its $P$-wave character relative to the ground-state atom is reflected by the node at the potential minimum, where the gradient of the wave function is maximal. The electron density is shown in the inset and resembles a $p$ state with the ground-state atom trapped between two maxima of the electron density.

The potential well around $R_\text{in}$ supports vibrational states which are visible in figure \ref{appfig:ULRMpot} as colored curves. The vibrational ground state is labeled $\chi_\text{in}(R)$ and serves as initial nuclear state.

Here, due to the smaller internuclear distance the addressable principal quantum numbers of the heavy Rydberg system are around $\nu^\ddagger=2172$, however, the laser wave length for the optical transition $\lambda^\ddagger = 2218.5\,\text{nm}$ is slightly larger, since the initial electronic state of the Rydberg molecule is lower in energy. The Franck-Condon overlap is $\braket{\chi_\text{in}}{\chi_{\nu^\ddagger}^0}=0.16$ and the energy spacing of the Rydberg series in this region is approximately $50\,\text{GHz}$.
The spatial contribution to the electronic transition
$\frac{\braket{\psi_\text{f}}{\vu{\epsilon}\v{r}|\psi_\text{in}}}{\braket{S}{\psi_\text{in}}}$
is in this case two orders of magnitude smaller than for the $24f$ state, but still on the same order of magnitude as the Rydberg transition from the ground state to the $12p$ state. \\
The spacial contribution of the electronic dipole transition can be increased by going to larger internuclear separations, however this results in a decrease of the spin contribution. For example, in the case of a $16p$ Rydberg molecule, the spacial contribution doubles, while the spin contribution halves leading to the same transition strength.

\section*{References}
\bibliographystyle{iopart-num}
\bibliography{hrs_24f.bbl}

\end{document}